\documentclass[10pt,journal,compsoc]{IEEEtran}
\ifCLASSOPTIONcompsoc
\usepackage[nocompress]{cite}
\else
\usepackage{cite}
\fi

\ifCLASSINFOpdf
\usepackage[pdftex]{graphicx}
\else
\fi

\usepackage{amsmath}

\usepackage{url}

\usepackage{amsfonts} 
\usepackage{pgfplots}
\usepackage{subcaption} 

\usepackage{multirow}

\usepackage[linesnumbered,ruled]{algorithm2e}

\newcommand\blfootnote[1]{%
	\begingroup
	\renewcommand\thefootnote{}\footnote{#1}%
	\addtocounter{footnote}{-1}%
	\endgroup
} 

\begin{document}
	
	\title{Privacy Preserving $n$-Party Scalar Product Protocol}
	\author{Florian~van Daalen,~\IEEEmembership{Graduate Student Member,~IEEE,}
		Lianne~Ippel,
		Andre~Dekker,
		and~ Inigo~Bermejo% <-this % stops a space
		\IEEEcompsocitemizethanks{\IEEEcompsocthanksitem  F. van Daalen, I. Bermejo, and A. Dekker are with 	the Department of Radiation Oncology (MAASTRO) GROW School for Oncology and Reproduction Maastricht University Medical Centre+ Maastricht the Netherlands\protect\\
			\IEEEcompsocthanksitem L. Ippel is with Statistics Netherlands Heerlen the Netherlands.}% <-this % stops a space
	}
	
	
	\IEEEtitleabstractindextext{%
		\begin{abstract}
			Privacy-preserving machine learning enables the training of models on decentralized datasets without the need to reveal the information, both on horizontally and vertically partitioned data. However, it requires specialized techniques and algorithms to perform the necessary computations. The privacy preserving scalar product protocol, which enables the dot product of vectors without revealing them, is one popular example for its versatility. For example it can be used to perform analyses that require counting the number of samples which fulfill certain criteria defined across various sites, such as calculating the information gain at a node in a decision tree. Unfortunately, the solutions currently proposed in the literature focus on two-party scenarios, even though scenarios with a higher number of data parties are becoming more relevant. In this paper, we propose a generalization of the protocol for an arbitrary number of parties, based on an existing two-party method. Our proposed solution relies on a recursive resolution of smaller scalar products. After describing our proposed method, we discuss potential scalability issues. Finally, we describe the privacy guarantees and identify any concerns, as well as comparing the proposed method to the original solution in this aspect. Additionally we provide an online repository containing the code.
		\end{abstract}
		
		\begin{IEEEkeywords}
			Federated Learning, $n$-party scalar product protocol, privacy preserving.
	\end{IEEEkeywords}}

	%\IEEEpeerreviewmaketitle
	\maketitle

	% is not selected <OR> if conference mode is selected - because compsoc
	% conference papers position the abstract like regular (non-compsoc)
	\IEEEdisplaynontitleabstractindextext

	% \fi
	
	\ifCLASSOPTIONcompsoc
	\IEEEraisesectionheading{\section{Introduction}\label{sec:introduction}}
	\else
	\section{Introduction \& related work}
	\label{sec:introduction}
	\fi

	\blfootnote{\emph{The views expressed in this paper are those of the authors and do not necessarily reflect the policy of Statistics Netherlands.}}
	
	\IEEEPARstart{F}{ederated} learning is a field that has recently grown in prominence due to increasing awareness of data privacy issues and data ownership as well as the rising need to combine data originating from different sources \cite{li_review_2020}. It is a thriving research field that promises to make it possible to apply machine learning algorithms (or any other data analysis) on multiple decentralized datasets in a collaborative manner \cite{kairouz_advances_2019}. This applies to both horizontally and vertically split data. Horizontally partitioned data describes the situation where different organizations collect the same information from different individuals (e.g. the same clinical data collected in multiple hospitals). Vertically partitioned data occurs when different organizations collect different information about the same individuals (e.g. insurance claims and hospital records).
	
	In order to apply machine learning algorithms on decentralized data, various techniques have been proposed to run the necessary analyses in a privacy-preserving manner. The techniques for vertically partitioned data are generally referred to with the umbrella term of secure multiparty computation (SMPC) \cite{yao_protocols_1982}. SMPC is a research field that focuses on developing methods to calculate functions on decentralized data without revealing the data to other parties.
	
	Examples of the various proposed techniques are machine learning algorithms to train Bayesian networks \cite{chen_fedbe_2021}, neural networks \cite{duan_self-balancing_2021}, or random forests \cite{liu_federated_2020}. These algorithms may rely on techniques such as secret sharing \cite{beimel_secret-sharing_2011} and homomorphic encryption \cite{parmar_survey_2014}. Both secret sharing and homomorphic encryption work at their core by transforming the original values  $\alpha$ and  $\beta$, owned by different parties, into transformed values  $\gamma$ and  $\delta$ such that $f( \alpha, \beta) = g( \gamma, \delta)$, thus making it possible to calculate the result of function $f( \alpha, \beta)$ by calculating a different function $g( \gamma, \delta)$ without ever needing to reveal  $\alpha$ or  $\beta$. In the case of homomorphic encryption, this is achieved by using encryption schemes that are ‘homomorphic’ with respect to specific functions, allowing the user to calculate these functions using encrypted data\cite{parmar_survey_2014}. In the case of secret sharing, the core concept relies on obfuscating the raw data with a secret share (e.g., a random number), and then applying calculations to the obfuscated data in such a way that the secret shares will cancel out in the end\cite{beimel_secret-sharing_2011}.
	
	Other techniques focus on specific calculations that can be used as building blocks for machine learning algorithms, such as the scalar product (or dot product) of vectors. The scalar product is an integral part of various machine learning algorithms, such as neural network training \cite{wiedemann_compact_2020}. Therefore, secure scalar product protocols have been widely studied in federated learning \cite{du_building_2002}. In addition, it can be used in combination with clever data representations to calculate various statistical measures in a privacy preserving manner, such as the information gain of an attribute, as well as to classify an individual using a decision tree in a federated setting \cite{du_building_2002}. More generally speaking the scalar product protocol can be employed to determine the size of a subset of the population that fulfills a set of criteria in a privacy preserving manner, even if the relevant attributes are spread across multiple data owners.
	
	Because of its importance, multiple scalar product variants have been proposed. Du and Atallah proposed several methods for the scalar product \cite{du_privacy-preserving_2001, goos_secure_2001}. Du et al. also proposed a similar method for secure matrix multiplication to be used in multivariate statistical analysis\cite{du_privacy-preserving_2004}. Vaidya and Clifton \cite{vaidya_privacy_2002} proposed a new method to alleviate the scalability issues of existing methods and used this method to determine globally valid association rules. Du and Zhan \cite{du_building_2002} proposed yet another alternative, with better time complexity than the method proposed by Vaidya and Clifton\cite{vaidya_privacy_2002}, and better communication cost than the methods proposed by Du and Atallah\cite{du_privacy-preserving_2001, goos_secure_2001}. Du and Zhan \cite{du_building_2002} then used it to train a decision tree in a federated setting. Goethals et al. \cite{hutchison_private_2005} discovered certain privacy flaws in some of the earlier mentioned protocols, and suggested an alternative with improved privacy guarantees. Shmueli and Tassa utilize a scalar product protocol to solve a problem with $n$ parties\cite{shmueli_mediated_2020}, however, it should be noted that they solely use the scalar product protocol to solve multiple independent $2$-party sub-problems. 
	
	However, all these solutions focus on two party scenarios where the scalar product is concerned. Translating them to scenarios involving more than two parties is not straightforward, if at all possible. This is a significant drawback since in practice often three, or even more parties, can be involved.
	
	In this study, we look at the method proposed by \cite{du_building_2002} and determine if, and how, it can be scaled to an arbitrary number of parties. This has applications for the various calculations which can (partially) be transformed into a scalar product problem mentioned before, such as calculating information gain or anything else that can be represented as a set-inclusion problem.
	
	\section{Method}
	In this section, we first introduce the notation used, then we describe the original solution proposed \cite{du_building_2002}. We will then try to naïvely translate the original solution to an $n$-party situation. This naïve translation will result in several left-over terms in the equations which need to be solved. We will then discuss how these left-over terms can be solved. We will illustrate the steps in this translation with a three-party scenario.  Finally, we will give a formal definition for the $n$-party scenario.
	
	In this paper, we use lowercase letters to denote scalars (e.g., ‘$s$’), uppercase for vectors (e.g., $V$) and uppercase with a bold face for matrices (e.g., ‘$\bold{M}$’).
	
	\subsection{Original protocol}
	The original protocol\cite{du_building_2002} works as follows. Alice and Bob have different features on the same individuals and want to calculate the scalar product of their private vectors $A$ and $B$, both of size $m$ where $m$ is semi-honest commodity server we have named Merlin. The protocol consists of the following steps.
	\begin{enumerate}
		\item Merlin generates two random vectors $R_{a}$, $R_{b}$ of size $m$ and two scalars $r_{a}$ and $r_{b}$ such that $r_{a} + r_{b} = R_{a} \cdot R_{b}$, where either $r_{a}$ or $r_{b}$ is randomly generated. Merlin then sends $\{R_{a}, r_{a}\}$ to Alice and $\{R_{b}, r_{b}\}$ to Bob.
		\item Alice sends $\hat{A} = A + R_{a}$ to Bob, and Bob sends $\hat{B} = B + R_{b}$ to Alice.
		\item Bob generates a random number $v_{2}$ and computes $u = \hat{A} \cdot B + r_{b} - v_{2}$, then sends the result to Alice.
		\item Alice computes $u - (R_{a} \cdot   \hat{B}) + r_{a} = A \cdot B - v_{2} = v1$ and sends the result to Bob.
		\item Bob then calculates the final result $v1 +v_{2} = A \cdot B$. 
	\end{enumerate}
	
	It should be noted that this protocol utilizes a secret sharing approach. Because of this, the extended $n$-party protocol will utilize the same secret sharing approach.
	
	\subsection{Naïve translation to a three-party scenario}
	For our three-party scenario we now have Alice, Bob and Claire who want to calculate the scalar product of their three vectors $A$, $B$, and $C$ of size $m$ as well as Merlin who will aid them in the calculation by fulfilling the role of commodity server. The first problem we encounter here is that $A \cdot B \cdot C$ does not result in a scalar, it results in another vector. This means it is impossible to simply chain the scalar product protocol. Hence, we must first translate our scalar product problem into a different form so it can be solved for multiple parties. 
	
	To do this we create three diagonal matrices, matrices where only the diagonal has non-zero values, $\bold{A}$, $\bold{B}$, and $\bold{C}$ of size $m \times m$, using the original vectors to fill the diagonals. This allows us to calculate $\bold{A} \cdot \bold{B} \cdot \bold{C}$, the result of which is a matrix. To turn this back into a scalar we define a function $\varphi$ which allows us to calculate the sum of the diagonal of a matrix. This means we have translated our $2$-party scalar product problem into a $3$-party matrix product problem where we calculate $\varphi(\bold{A} \cdot \bold{B} \cdot \bold{C})$. This naïve translation has a similar form as the matrix multiplication method proposed by Du et al. \cite{du_privacy-preserving_2004} mentioned earlier in this article, however, it includes more than two parties and all of our matrices are diagonal matrices. 
	
	It should be noted that this matrix multiplication method cannot simply be used to replace the scalar product protocol, as this would result in individual level data being shared across parties. For example, when using the scalar product protocol to build a decision tree\cite{du_building_2002}, we have diagonal matrices, and the diagonal only contains 0 and 1 values. It would be trivial to deduce which positions only contained a value of 1 at all parties based on the final result using the matrix multiplication approach, which would be a major breach of privacy, as this would allow one to know which individuals were selected.
	
	Having successfully translated our problem into a form where we can work with three parties, we will now attempt to naïvely translate the protocol. First, it should be noted that Merlin should generate random diagonal matrices instead of vectors. Second, he needs to generate an extra matrix $\bold{R_{c}}$ and scalar $r_{c}$ to send to Claire. Third, we need to introduce an extra step into our protocol for Claire that is equivalent to step 4 in the two-party protocol. And last, wherever vectors owned by Alice and Bob are multiplied we must now multiply matrices owned by Alice, Bob and Claire. It should also be noted that whenever we are now multiplying matrices, we need to apply the $\varphi$ function to turn the resulting matrix into a scalar. Consequently, our naïvely adapted protocol will look as follows:
	
	\begin{enumerate}
		\item Merlin generates three random diagonal matrices $\bold{R_{a}}$, $\bold{R_{b}}$, $\bold{R_{c}}$ and two random scalars $r_{a}, r_{b}$. It then calculates a third scalar $r_{c}$ such that $r_{a} + r_{b} + r_{c} = \varphi(\bold{R_{a}} \cdot \bold{R_{b}} \cdot \bold{R_{c}})$. Merlin then sends $\{\bold{R_{a}}, r_{a}\}$ to Alice, $\{\bold{R_{b}}, r_{b}\}$ to Bob and $\{\bold{R_{c}}, r_{c}\}$ to Claire.
		\item Alice calculates   $\bold{\hat{A}} = \bold{A} + \bold{R_{a}}$ and sends it to Bob and Claire, Bob sends   $\bold{\hat{B}} = \bold{B} + \bold{R_{b}}$ to Alice and Claire, and Claire sends $\bold{\hat{C}} = \bold{C} + \bold{R_{c}}$ to Alice and Bob.
		\item Bob generates a random number $v_{2}$ and computes $u_{1} = \varphi(  \bold{\hat{A}} \cdot \bold{\hat{C}} \cdot \bold{B} ) + r_{b} - v_{2}$, then sends the result to Alice.
		\item Alice computes $u_{2}= u_{1}  - \varphi(\bold{R_{a}} \cdot   \bold{\hat{B}} \cdot \bold{\hat{C}}) + r_{a}$, then sends the result to Claire
		\item Claire then computes $u_{3}  = u_{2} - \varphi( \bold{R_{c}} \cdot \bold{\hat{A}} \cdot   \bold{\hat{B}})+ r_{c}$.  Claire then sends $u_{3}$ to Bob.
		\item Bob then calculates the final result $ u_{3} + v_{2} = \varphi(\bold{A} \cdot \bold{B} \cdot \bold{C} ) - \varphi(\bold{R_{a}} \cdot \bold{R_{b}} \cdot \bold{R_{c}}) - \varphi(\bold{A} \cdot \bold{R_{b}} \cdot \bold{R_{c}}) - \varphi(\bold{B} \cdot \bold{R_{a}} \cdot \bold{R_{c}}) - \varphi(\bold{C} \cdot \bold{R_{a}} \cdot \bold{R_{b}})$\footnote{A full elaboration of the equation can be found in appendix \ref{Full 3-party naïve calculation}}
	\end{enumerate}
	
	As we can see our final result is not equal to $\varphi(\bold{A} \cdot \bold{B} \cdot \bold{C})$ because there are several left-over terms (i.e., $\varphi(\bold{R_{a}} \cdot \bold{R_{b}} \cdot \bold{R_{c}})$, $\varphi(\bold{A} \cdot \bold{R_{b}} \cdot \bold{R_{c}})$, $\varphi(\bold{B} \cdot \bold{R_{a}} \cdot \bold{R_{c}})$, and $\varphi(\bold{C} \cdot \bold{R_{a}} \cdot \bold{R_{b}}))$.

	\subsection{Solving the left-over terms}
	The first left-over that should be solved is the left-over of the form $\varphi(\bold{R_{a}} \cdot \bold{R_{b}} \cdot \bold{R_{c}})$. The protocol will naturally result in a left-over term of the form $(n-2) \varphi( \bold{R_{a}} \cdot \bold{R_{b}} \cdot \bold{R_{c}})$ because we already add the various $r_{x}$ for each $x \in \{a,b,c\}$ once in step $3-5$, even in the naïve translation. We can solve this leftover term simply by replacing $r_{x}$ in step $3-5$ with $(n-1)r_{x}$, because $(n-1) \varphi(\bold{R_{a}} \cdot \bold{R_{b}} \cdot \bold{R_{c}}) = (n-1) (r_{a} + r_{b} + r_{c})$. For example in step 4 instead of adding $r_{a}$ we will add $2r_{a}$ in the $3$-party protocol.
	
	The remaining left-over terms are $\varphi(\bold{A} \cdot \bold{R_{b}} \cdot \bold{R_{c}})$,  $\varphi(\bold{B} \cdot \bold{R_{a}} \cdot \bold{R_{c}})$, and $\varphi(\bold{C} \cdot \bold{R_{a}} \cdot \bold{R_{b}})$. These left-over terms all have the form of $\varphi(\bold{X} \cdot \bold{R_{y}} \cdot \bold{R_{z}})$, where $x, y,$ \& $z$ represent the different parties Alice, Bob, \& Claire, and each of the multiplicands always belongs to a different party (e.g., they are never of the form $\varphi(\bold{X} \cdot \bold{R_{x}} \cdot \bold{R_{y}})$). Furthermore, the combined term $\bold{R_{y}} \cdot \bold{R_{z}}$ is known by Merlin, hence this can be rewritten as $\varphi(\bold{X} \cdot \bold{M})$, where $\bold{M} = \bold{R_{y}} \cdot \bold{R_{z}}$ and is owned by Merlin. This means that this left-over problem can be simplified into a $2$-party scalar product problem, where Merlin is one of the parties.  More generally these left-over terms within an $n$-scalar product protocol are themselves $n-1$, or smaller, scalar product problems. These smaller scalar product protocols need to be solved with additional commodity servers (i.e., Merlin cannot play that role because he is involved as a party). In section \ref{scalability} we will discuss how many commodity servers are needed for a given $n$-party protocol.
	
	With the left-over terms solved we can now create a fully translated protocol to our three-party scenario.
	
	\subsection{Correct adaptation to a three-party scenario}
	To allow Alice, Bob, and Claire to calculate $\varphi(\bold{A} \cdot \bold{B} \cdot \bold{C})$ the following protocol should be followed.
	
	\begin{enumerate}
		\item Merlin generates three random diagonal matrices $\bold{R_{a}}$, $\bold{R_{b}}$, $\bold{R_{c}}$ and two random scalars $r_{a}, r_{b}$. It then calculates a third scalar $r_{c}$ such that $r_{a} + r_{b} + r_{c} = \varphi(\bold{R_{a}} \cdot \bold{R_{b}} \cdot \bold{R_{c}})$. Merlin then sends $\{\bold{R_{a}}, r_{a}\}$ to Alice, $\{\bold{R_{b}}, r_{b}\}$ to Bob and $\{\bold{R_{c}}, r_{c}\}$ to Claire.
		\item Alice sends   $\bold{\hat{A}} = \bold{A} + \bold{R_{a}}$ to Bob and Claire, Bob sends   $\bold{\hat{B}} = \bold{B} + \bold{R_{b}}$ to Alice and Claire, and Claire sends $\bold{\hat{C}} = \bold{C} + \bold{R_{c}}$ to Alice and Bob.
		\item Bob generates a random number $v_{2}$ and computes $u_{1} = \varphi(  \bold{\hat{A}} \cdot \bold{\hat{C}}\cdot \bold{B}) + 2r_{b} - v_{2}$, then sends the result to Alice.
		\item Alice computes $u_{2}= u_{1}  - \varphi(\bold{R_{a}} \cdot   \bold{\hat{B}} \cdot \bold{\hat{C}}) + 2r_{a}$, then sends the result to Claire
		\item Claire then computes $u_{3}=u_{2} -\varphi(\bold{R_{c}} \cdot \bold{\hat{A}} \cdot   \bold{\hat{B}}) + 2r_{c} = \varphi(\bold{A} \cdot \bold{B} \cdot \bold{C})- \varphi(\bold{A} \cdot \bold{R_{b}} \cdot \bold{R_{c}}) - \varphi(\bold{B} \cdot \bold{R_{a}} \cdot \bold{R_{c}}) - \varphi(\bold{C} \cdot \bold{R_{a}} \cdot \bold{R_{b}}) - v_{2}$
		\item The left-over terms $\varphi(\bold{A} \cdot \bold{R_{b}} \cdot \bold{R_{c}})$,  $\varphi(\bold{B} \cdot \bold{R_{a}} \cdot \bold{R_{c}})$, and  $\varphi(\bold{C} \cdot \bold{R_{a}} \cdot \bold{R_{b}})$ are solved by separate two-party scalar product protocols. The results are given to Claire and she computes $\varphi(\bold{A} \cdot \bold{B} \cdot \bold{C})- \varphi(\bold{A} \cdot \bold{R_{b}} \cdot \bold{R_{c}}) - \varphi(\bold{B} \cdot \bold{R_{a}} \cdot \bold{R_{c}}) - \varphi(\bold{C} \cdot \bold{R_{a}} \cdot \bold{R_{b}} ) + \varphi(\bold{A} \cdot \bold{R_{b}} \cdot \bold{R_{c}}) + \varphi(\bold{B} \cdot \bold{R_{a}} \cdot \bold{R_{c}}) + \varphi(\bold{C} \cdot \bold{R_{a}} \cdot \bold{R_{b}}) - v_{2} = \varphi(\bold{A} \cdot \bold{B} \cdot \bold{C}) - v_{2} = u_{3}$. Claire then sends $u_{3}$ to Bob.
		\item Bob then calculates the final result: $v_{2} + u_{3} = \varphi(\bold{A} \cdot \bold{B} \cdot \bold{C})$
	\end{enumerate}
	
	We have now successfully translated the two-party scalar product protocol into a three-party protocol.\footnote{A practical example of a $3$-party scalar product protocol can be found in appendix \ref{Full 3-party example}}
	
	\subsection{Full translation to an $n$-party scenario}
	\label{full translation}
	The $n$-party protocol can be formalized as follows:
	
	\begin{enumerate}
		\item If $n = 2$, use the two-party protocol \cite{du_building_2002}, else go to next step.
		\item Let $\bold{D_{1}}, \bold{D_{2}},…, \bold{D_{n}}$ be the diagonal matrices containing the vectors owned by the $n$ parties.
		\item Let $\varphi$ be a function that calculates the sum of the diagonal of a matrix.
		\item $\bold{R_{1}}, \bold{R_{2}},.., \bold{R_{n}}$ are random diagonal matrices generated by a commodity server Merlin.
		\item Let $\varphi(\bold{R_{1}} \cdot \bold{R_{2}} \cdot..\cdot \bold{R_{n}}) = r_{1} + r_{2} + … + r_{n}$ where all but one of the $r_{i}$ terms are randomly generated.
		\item Merlin shares the pairs $\{\bold{R_{i}},r_{i}\}$ with the $i$’th party for each $i \in [1,n]$
		\item All parties calculate $\bold{\hat{D_{i}}} = \bold{D_{i}} + \bold{R_{i}}$ and share the result
		\item Party 1 generates $v_{2}$.
		\item Party 1 then calculates $u_{1} =  \varphi(\prod_{i=2}^{n}\bold{\hat{D}_{i}} \cdot \bold{D_{1}})+ (n-1)\cdot r_{1}- v_{2}$
		\item For each other party $i$ calculate $u_{i} = u_{i-1} - \varphi((\prod_{x=1}^n {\bf \hat{D}_x}|x\neq i)\cdot {\bf R_i}) + (n-1) \cdot r_{i}$
		\item This results in $\varphi(\bold{D_{1}} \cdot \bold{D_{2}} \cdot .. \cdot \bold{D_{n}}) - \bold{L_{1}} - \bold{L_{2}} -.. \bold{L_{n}} - v_{2}$ Where $\bold{L_{i}}$ corresponds to leftover terms of the form $\varphi(\prod_{i=1}^{m} \bold{D_{i}} \prod_{j=m}^{n} \bold{R_{j}}$|$i \neq j)$, where all parties are involved, either as ${\bf D_i}$, providing their raw data, or as ${\bf R_j}$, using their random matrix, but never as both.		
		\item These leftover terms represent a scalar product problem of at most $n-1$ parties. Thus these sub problems can be solved separately using a smaller $n$-party scalar product protocol.
		\item Solving these leftover terms allows party $n$ to calculate $\varphi(\bold{D_{1}} \cdot \bold{D_{2}} \cdot..\cdot \bold{D_{n}}) - v_{2} = u_{n}$
		\item Party 1 can then calculate the final result $u_n + v_{2} = \varphi(\bold{D_{1}} \cdot \bold{D_{2}} \cdot..\cdot \bold{D_{n}})$	
	\end{enumerate}
	
	This allows us to calculate the scalar product for an arbitrary amount of parties. Pseudocode of the protocol can be found in algorithm \ref{pseudocode}. Now that we have shown that the protocol can be translated to a scenario with arbitrary $n$ we will discuss how the protocol scales as well as potential security issues in the next section.
	
	\begin{algorithm}
		\SetKwInOut{Input}{Input}
		\SetKwInOut{Output}{Output}	
		\SetKw{KwBy}{by}
		
		\underline{nPartyScalarProduct($\mathcal{D}$)} \\
		\Input{ The set $\mathcal{D}$ of diagonal matrices $\bold{D_{1}} .. \bold{D_{n}}$ containing the original vectors owned by the $n$ parties}
		\Output{$\varphi(\bold{D_{1}} \cdot \bold{D_{2}} \cdot..\cdot \bold{D_{n}})$}
		\eIf{$|\mathcal{D}| = 2$}
		{
			return $2$-party scalar product protocol($\mathcal{D}$);
		}
		{		
			\For{$i\gets0$ \KwTo $|\mathcal{D}|$ \KwBy $1$}{
				$\bold{R_{i}} \gets generateRandomDiagonalMatrix()$ \\
			}
			Let $\varphi(\bold{R_{1}} \cdot \bold{R_{2}} \cdot..\cdot \bold{R_{n}}) = r_{1} + r_{2} + … + r_{n}$ \\
			Share $\{\bold{R_{i}},r_{i}\}$ with the $i$'th party for each $i \in [1,n]$ \\
			$v_{2} \gets randomInt()$\\
			$u_1 \gets  \varphi(\prod_{i=2}^{n}\bold{\hat{D}_{i}} \cdot \bold{D_{1}})+ (n-1)\cdot r_{1}- v_{2}$ \\
			\For{$i\gets2$ \KwTo $|\mathcal{D}|$ \KwBy $1$}{
				$u_{i} = u_{i-1} - \newline \varphi((\prod_{x=1}^n {\bf \hat{D}_x}|x\neq i)\cdot {\bf R_i}) \newline + (n-1) \cdot r_{i}$
			}
			$y \gets u_{n}$ \\
			\For{subprotocol $\in$ determineSubprotocols($\mathcal{D}, \mathcal{R}$)}{
				$y \gets y -$ nPartyScalarProduct($subprotocol$)
			}
			return $y + v_{2}$		
		}
		\underline{determineSubprotocols($\mathcal{D}, \mathcal{R}$)} \\
		\Input{The set $\mathcal{D}$ of diagonal matrices $\bold{D_{1}} .. \bold{D_{n}}$ of the original protocol. The set $\mathcal{R}$ of random diagonal matrices used in the original protocol}
		\Output{The sets $\mathcal{D}_{subprotocol}$ for each subprotocol}
		
		\For{$k\gets2$ \KwTo $|\mathcal{D}| -1$ \KwBy $1$}{
			$uniqueCombinations \gets selectKSizedCombosFromSet(k, \mathcal{D})$ \\
			\For{$selected \in uniqueCombinations$}{
				$subprotocol \gets  \bold{D_{i}} | i \in selected  + \bold{R_{j}} | j \not \in selected$
				$\mathcal{D}_{subprotocols} \gets  \mathcal{D}_{subprotocols} +  subprotocol$ 	
			}
		}
		
		return 	$\mathcal{D}_{subprotocols}$
		\caption{The n-party scalar product protocol}\label{pseudocode}
	\end{algorithm}
	
	\subsection{Commodity server}
	The $n$-party scalar product protocol contains multiple sub-protocols of at most $n-1$ sized all of which involve data owned by the commodity server in the $n$-party scalar protocol. These sub protocols will need to use a commodity server as well. However, the original commodity server Merlin cannot be reused as Merlin fulfills the role of data-owner in these sub protocols. In section \ref{scalability} we will discuss what influence this will have as $n$ grows and how potential issues can be minimized.
	
	\section{Discussion}
	In this paper, we have translated an existing $2$-party scalar product \cite{du_building_2002} protocol to an $n$-party protocol. We have shown that a naïve translation is insufficient. However, by using a more sophisticated approach, it is possible to adapt the protocol to work with an arbitrary number of parties.  In appendix \ref{Full 3-party example}, a fully worked out example of the three-party protocol can be found. Appendix \ref{git} provides references to a repository containing java and python implementations of the $n$-party protocol.
	
	We will now discuss the security and privacy guarantees this $n$-party protocol provides as well as how the complexity scales as the number of parties grows and how practical it is to use this protocol.
	
	\subsection{Security}
	\label{security}
	The proposed method requires a commodity server, which is a semi-honest trusted third party within the calculation. A semi-honest party is a party which executes its part in the protocol accurately, but may try to learn as much as it can from the messages it receives in the process \cite{do_role_2019}. In this section we will discuss the exact risks involved with this.
	
	As a method that relies on secret shares generated by a semi-trusted third party, this protocol utilizes an approach similar to assymetric encryption\cite{schneider_comment_2020}, with the individual secret shares performing the role of private keys. This limits the risks involved. However, the trusted third party does introduce a risk in itself.
	
	The risk posed by requiring a semi-honest trusted third party to be the commodity server would be that several semi-honest parties could potentially cooperate with the commodity server in order to jointly learn private data of the other parties. It should be noted that this risk is higher in an Internet of Things (IoT) setting than in a formalized joint research setting. An IoT setting consists of many unverified devices and parties. A formal joint research setting allows all parties involved to verify, and enforce, for example by requiring audits and adding other legal agreements, the integrity of the other parties to a certain extent. This will minimize the risk in practice in this setting. While it would be preferable if privacy could be protected by design with technical solutions, there will always be a need for a certain degree of trust in the various parties involved and legal means are a perfectly acceptable way of achieving the required trust \cite{kairouz_advances_2019}.
	
	However, this does not remove the technical possibility of a joint attack when all parties are semi-honest. The local calculations done at a given node $i$ are always of the form: $u_{i} = u_{i-1} - \varphi((\prod_{x=1}^{n} \bold{\hat{D}_{x}} |x \neq i ) \cdot \bold{R_{i}}) + (n-1) \cdot r_{i}$. Where $\bold{\hat{D_{x}}}$ is locally known by every data-owner participating in this protocol. However, $\bold{\hat{D_{x}}}$ is unknown to the commodity server in this protocol. Assuming the node cooperates with the commodity server, they could then separate $\bold{\hat{D_{x}}}$ into its components $\bold{D_{x}}$ and $\bold{R_{x}}$. Where $\bold{D_{x}}$ is private data belonging to a different party and $\bold{R_{x}}$ is the random diagonal matrix generated by the commodity server, thus learning $\bold{D_{x}}$. This is a serious concern. This issue is especially relevant in an IoT setting where the trustworthiness of the commodity servers and individual parties is very difficult to verify and enforce.
	
	However, in a formal joint research setting, a sufficient level of trust can be achieved to minimize the risk of this attack by enforcing the commodity server to act as an honest party, not just semi-honest\cite{kairouz_advances_2019}\cite{truong_privacy_2021} First, it is possible to simply enforce this using legal means and mandate it is honest, however this may not be accepted in practice. Second, it is possible to give all parties involved joint custody over the commodity servers, thus allowing each party to individually verify the commodity server is completely honest.
	
	Joint custody over the commodity servers could, for example, be achieved by allowing any party to execute independent audits of the commodity server and giving them a veto over the hardware and software setup used on the servers. Such a setup allows each party to individually verify that the commodity server is honest, which works because each party has a vested interested in ensuring the honesty of the commodity server to protect their own data. This should allow the parties to jointly guarantee the commodity server are honest, even if the individual parties themselves are semi-honest.
	
	It is important to note that these security concerns, and the possible solutions, are the same regardless of the size of $n$. That is to say, our proposed $n$-party protocol is equally as secure as the original $2$-party protocol proposed by Du and Zhan because the original protocol also uses a trusted third party as commodity server which as we just discussed is the vulnerability exploited in a collusion attack.

	\subsection{Scalability}\label{scalability}
	The number of subprotocols will grow with a factorial order of growth with respect to $n$. The reason it scales in this manner is because the subprotocols have the form of $\varphi(\prod_{i=1}^{m} \bold{D_{i}} \prod_{j=m}^{n} \bold{R_{j}}$|$i \neq j)$. Where all parties are involved, either as $\bold{D_{i}}$, providing their raw data, or as $\bold{R_{j}}$, using their random matrix, but never as both. There will be $\frac{n!}{x!(n-x)!}$ such subprotocols for each $2 \leq x < n$.
	
	These subprotocols will have $x$ $\bold{R_{j}}$ factors and $n-x$ $\bold{D_{i}}$ factors. For example, a three-party protocol will have the following $3$ subprotocols involving $2$ $\bold{R_{j}}$ factors: $\varphi(\bold{A} \cdot \bold{R_{b}} \cdot \bold{R_{c}})$, $\varphi( \bold{B} \cdot \bold{R_{a}} \cdot \bold{R_{c}})$ and $\varphi(\bold{C} \cdot \bold{R_{a}} \cdot \bold{R_{b}})$. A $4$ party protocol will have $4$ subprotocols involving $3$ $\bold{R_{j}}$ factors: $2 \cdot \varphi(\bold{A} \cdot \bold{R_{b}} \cdot \bold{R_{c}} \cdot \bold{R_{d}})$, $2\cdot \varphi(\bold{B} \cdot \bold{R_{a}} \cdot \bold{R_{c}} \cdot \bold{R_{d}})$, $2\cdot \varphi(\bold{C}\cdot \bold{R_{a}}\cdot \bold{R_{b}} \cdot \bold{R_{d}})$, and $2\cdot \varphi(\bold{D} \cdot \bold{R_{a}}\cdot \bold{R_{b}} \cdot \bold{R_{c}})$. As well as $6$ subprotocols involving $2$ $R_{j}$ factors: $\varphi(\bold{A} \cdot \bold{B} \cdot \bold{R_{c}} \cdot \bold{R_{d}})$, $\varphi(\bold{A} \cdot \bold{C} \cdot \bold{R_{b}} \cdot \bold{R_{d}})$, $\varphi({\bf A}\cdot {\bf D}\cdot {\bf R_b}\cdot {\bf R_c})$, $\varphi(\bold{B}\cdot \bold{C} \cdot \bold{R_{a}} \cdot \bold{R_{d}})$, $\varphi(\bold{B}\cdot \bold{D}\cdot \bold{R_{a}} \cdot \bold{R_{c}})$ and $\varphi(\bold{C}\cdot \bold{D}\cdot \bold{R_{a}}\cdot \bold{R_{b}})$.
	
	This growth in subprotocols will have an effect on the scalability. We will discuss the two aspects in which this matters in the following two sections.
	
	\subsubsection{Time and Space Complexity}
	The first aspect affected by the factorial order of growth is the time complexity of the protocol. The amount of direct subprotocols for an $n$-party protocol will be equal to $\frac{n!}{x!(n-x)!}$ for each $x \in [2;n]$. These subprotocols may also have further subprotocols themselves. Furthermore, the amount of messages that need to be send for a given protocol are as follows; $1$ message needs to be send from the commodity server to each of the $n$ dataowners to share the relevant pair of $\{\bold{R_{i}}, r_{i}\}$. Each party then shares its matrix $\bold{\hat{D}_{i}}$ with each other party, resulting in $n\cdot(n-1)$ messages. Finally each party has to share its subresult once, resulting in a further $n$ messages. This means a total of $n+n^{2}$ messages for a given protocol.
	
\begin{figure}[h]
		\centering
		\begin{tikzpicture}	
			\begin{axis}[
				ymode=log,
				legend pos=south east,
				xlabel=Number of parties
				]
				\addplot[
				color=blue,
				mark=square,
				]
				coordinates {
					(2,1)(3,4)(4,29)(5,336)(6,5687)(7,132294)
					
				};
				\addplot[
				color=red,
				mark=triangle,
				]
				coordinates {
					(2,6)(3,30)(4,224)(5,2600)(6,44008)(7,1023736)
				};

				\addlegendentry{Protocols}
				\addlegendentry{Messages}
				
			\end{axis}
		\end{tikzpicture}		
		\caption{Rate at which the number of protocols and messages grow as functions of $n$. (y-axis in log-scale)}
		\label{protocols}
	\end{figure}
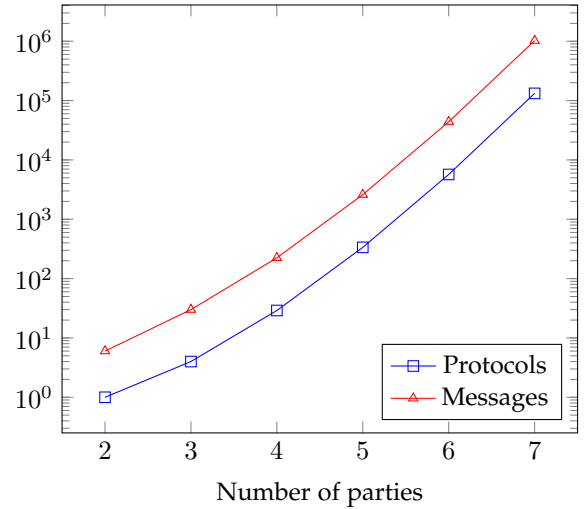

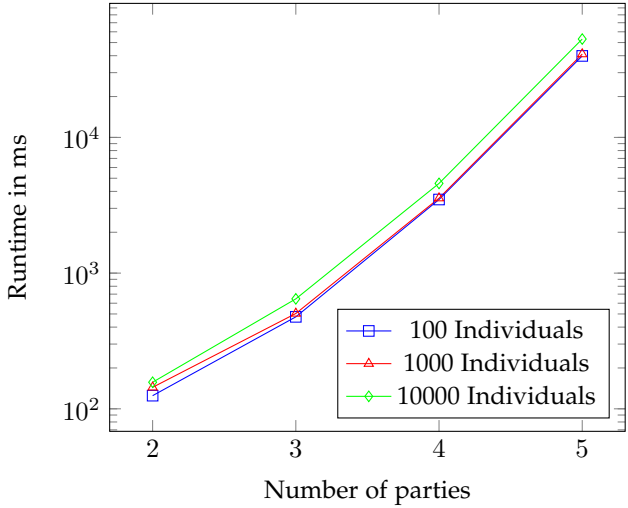
\begin{figure}[h]
		\centering
		\begin{tikzpicture}	
			\begin{axis}[
				ymode=log,
				legend pos=south east,
				xtick=data,
				xlabel=Number of parties,
				ylabel=Runtime in ms
				]
				\addplot[
				color=blue,
				mark=square,
				]
				coordinates {
					(2,125)(3,476)(4,3486)(5,39845)
					
				};
				\addplot[
				color=red,
				mark=triangle,
				]
				coordinates {
					(2,144)(3,504)(4,3566)(5,41101)
				};
				\addplot[
				color=green,
				mark=diamond,
				]
				coordinates {
					(2,157)(3,646)(4,4582)(5,53053)
				};

				\addlegendentry{100 Individuals}
				\addlegendentry{1000 Individuals}
				\addlegendentry{10000 Individuals}			
			\end{axis}
		\end{tikzpicture}		
		\caption{Average time in ms necessary to calculate the number of individuals fullfilling the requirements of 2-5 attributes divided over 2-5 parties for different population sizes. (y-axis in log-scale)}
		\label{runtime}
	\end{figure}

	In order to put this into perspective we show the number of protocols as a function of $n$ in figure \ref{protocols}. In addition to this, the results of a small experiment measuring the runtime performance, where the $n$-party protocol was used to calculate the number of individuals fullfilling certain attribute requirements, can be found in figure \ref{runtime}. This experiment was run on a windows laptop using an Intel(R) Core(TM) i7-10750H processor with 16GB of memory and 6 cores. All parties had a local datastation on this laptop, no significant optimization was implemented.
	
	As can be seen in figure \ref{protocols} the required number of protocols and messages grow quickly as $n$ grows. This is a significant downside of this protocol.  The results of the small runtime experiment further supports this, as the runtime does grow rapidly as the number of parties grows. However, it also shows that the protocol can easily deal with larger datasets as dataset size barely influences the runtime.
	It should also be noted that there is considerable room for parallelization within the protocol, allowing the protocol to still be useable in practice.
	The following steps can be parallelized: first, every subprotocol can naturally be calculated in parallel as these are independent problems. Secondly every calculation in substep $11$ detailed in section \ref{full translation} can be calculated in parallel as well. Both options will reduce the running time of the protocol, considerably, allowing it to still be a practical solution in many settings. In addition to this, the actual use of the protocol within model training can be optimized, for example by running multiple $n$-party product protocols in parallel.
	
	\subsubsection{Commodity Servers}
	It should be noted that these subprotocols need their own commodity server because no party may be both data owner and commodity server in a given protocol. Hence, we cannot reuse the original commodity server Merlin as it fullfills the role of a data owner in the subprotocols.
	
	A naïve solution to the problem posed by this need would be to set up sufficient commodity servers to deal with every sub-protocol. However, the amount of commodity servers needed will scale linearly with $n$, since a commodity server can be shared across all subprotocols of the same size. As the largest subprotocol in an $n$-party protocol will be an $(n-1)$-party subprotocol, and a two-party protocol will have no subprotocol, we will need $n-1$ commodity servers to solve an $n$-party problem. While this might be manageable for small $n$ this eventually becomes untenable. 
	
	An alternative to this naïve solution would be to have the various parties double as commodity servers whenever they are not involved in a calculation themselves. To show that this is a viable, and safe solution, we will first divide the subprotocols into two categories. All subprotocols have the form $\varphi(\prod_{i=1}^{m} \bold{D_{i}} \prod_{j=m}^{n} \bold{R_{j}} | i \neq j)$, this can be further subdivided into subprotocols which contain only $1$ $\bold{D_{i}}$ term, which will have the form $\varphi(\bold{D_{i}} \cdot \bold{R_{j}} \cdot... \cdot \bold{R_{m}})$,
	and subprotocols with multiple $\bold{D_{i}}$ terms.
	
	The first category of subprotocols, which only contain one $\bold{D_{i}}$  term, can be solved by simply sharing the result of random matrices $\bold{R_{j}} \cdot \bold{R_{j}}\cdot \cdot…\cdot \cdot \bold{R_{m}}$ with the owner of $\bold{D_{i}}$. $\bold{R_{j}} \cdot\bold{R_{j}}\cdot \cdot…\cdot \cdot \bold{R_{m}}$ is itself a random matrix, provided there are at least two $\bold{R_{j}}$ factors involved, which cannot be used to leak any information. For example, the sub-protocols in the three-party protocol can be solved this way without requiring extra commodity servers. Doing this will also be faster than using the two-party scalar product protocol as it only requires a straightforward multiplication instead of the entire scalar product protocol. It should however be noted that the solution to this subprotocol may never be revealed to the commodity server that owns the $\bold{R_{j}}$ terms, as this would allow the commodity server to calculate $\bold{D_{i}}$. For example, if we are calculating $\bold{A}\cdot\bold{R_{b}}\cdot\bold{R_{c}}$ the result should never be revealed to Merlin, as revealing this would allow Merlin to learn Alice's data. This is of course also true in the original $2$-party protocol.
	
	The second category of subprotocol, which contains multiple $\bold{D_{i}}$ terms, can reuse one of the parties which is not currently providing data (i.e. a $\bold{D_{i}}$ term) as the new commodity server. This is secure as there is no need to reveal anything to the commodity server during the calculation. All it needs to do is generate and share the new $\{{\bf R_i}, r_i\}$ pairs for this subprotocol. As such, it never needs to see any (sub)results, and thus cannot reverse engineer anything. Additionally,  the same party should never be used twice as a commodity server in any set of subprotocols. That is to say, if Alice handles a $3$-party subprotocol then she should not handle any child protocols that arise as a consequence of this specific $3$-party subprotocol. Fortunately, it is easy to avoid this as there will always be at least one new party available to fulfil the role of commodity server for the new subprotocols. 
	
	While this is a practical solution to the need for multiple commodity servers, it does come with the major caveat that one must be certain no parties will attempt to cooperate to jointly learn private data of the other parties. As pointed out in section \ref{security}, the protocol is vulnerable to this type of attack.
	
	\section{Conclusion}
	In this paper, we have explained how the two-party scalar product protocol by Du and Zhan \cite{du_building_2002} can be scaled to an $n$-party scalar product protocol. We have illustrated how it works using a three-party scenario, after which we have given the formal definition of the protocol for any number of parties. This protocol can be used to calculate a number of metrics, such as the information gain of an attribute \cite{du_building_2002}, in a scenario with an arbitrary number of parties. The benefit of being able to calculate such metrics is that it opens up the door for other more complex analysis. For example, using the information gain one can build a decision tree or apply feature selection. 
	
	Similarly, by using an innovative data representation the $n$-party protocol can be used to classify an individual in a privacy preserving manner using a decision tree \cite{du_building_2002}. By using other innovative data representations this $n$-party protocol could potentially be used for a wide variety of analysis and calculations. Aside from these benefits, which require the problem at hand to be rephrased into a scalar product problem, there is also the obvious benefit that it allows the use of the scalar product itself in an $n$-party scenario. This allows the use of any calculation that would normally rely on the scalar product in a classical machine learning setting but which cannot be executed easily in a federated setting without a private $n$-party scalar product protocol.
	
	While not appropriate in every scenario (scalability and the need for more commodity servers or semi-honest servers as the number of parties grows are a practical concern), we believe this is still a valuable tool in the federated learning toolbox.
	
	\subsection{Future work}
	For future work we would like to devise $n$-party protocols with better time complexity, as well as find a way to remove the vulnerability to joint-attacks introduced by the need for a commodity server.
	
	In addition to this it would be valuable to investigate to which extend our extension to $n$ parties can be applied to the secure matrix multiplication proposed by Du et al.\cite{du_privacy-preserving_2004}. The protocol used for matrix multiplication is very similar to the $2$-party scalar product protocol we extended, as such our extension should be of use when extending this matrix multiplication protocol.
	
	Lastly, we are planning to utilize the $n$-party scalar product protocol to implement various federated algorithms so we can test the practical viability of this protocol in a real life setting. 
	
	\bibliographystyle{IEEETran}
	\bibliography{nPartyProtocol}
		
	\begin{IEEEbiography}[{\includegraphics[width=1in,height=1.25in,clip,keepaspectratio]{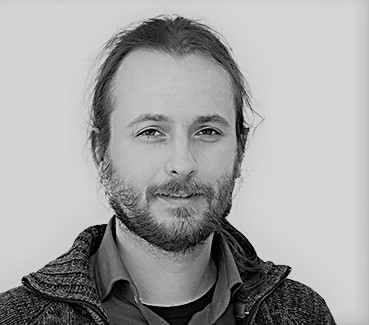}}]{Florian van Daalen}
		Florian van Daalen received his BSc degree in Knowledge Engineering from University Maastricht in 2012 and his MSc degree in Artificial Intelligence in 2014. He is currently working toward the PhD degree in Clinical Data Science within the Clinical Data Science group, University Maastricht, Netherlands. His research interests include privacy preserving techniques, federated learning, and ensemble based learning.
	\end{IEEEbiography}
	\begin{IEEEbiography}[{\includegraphics[width=1in,height=1.25in,clip,keepaspectratio]{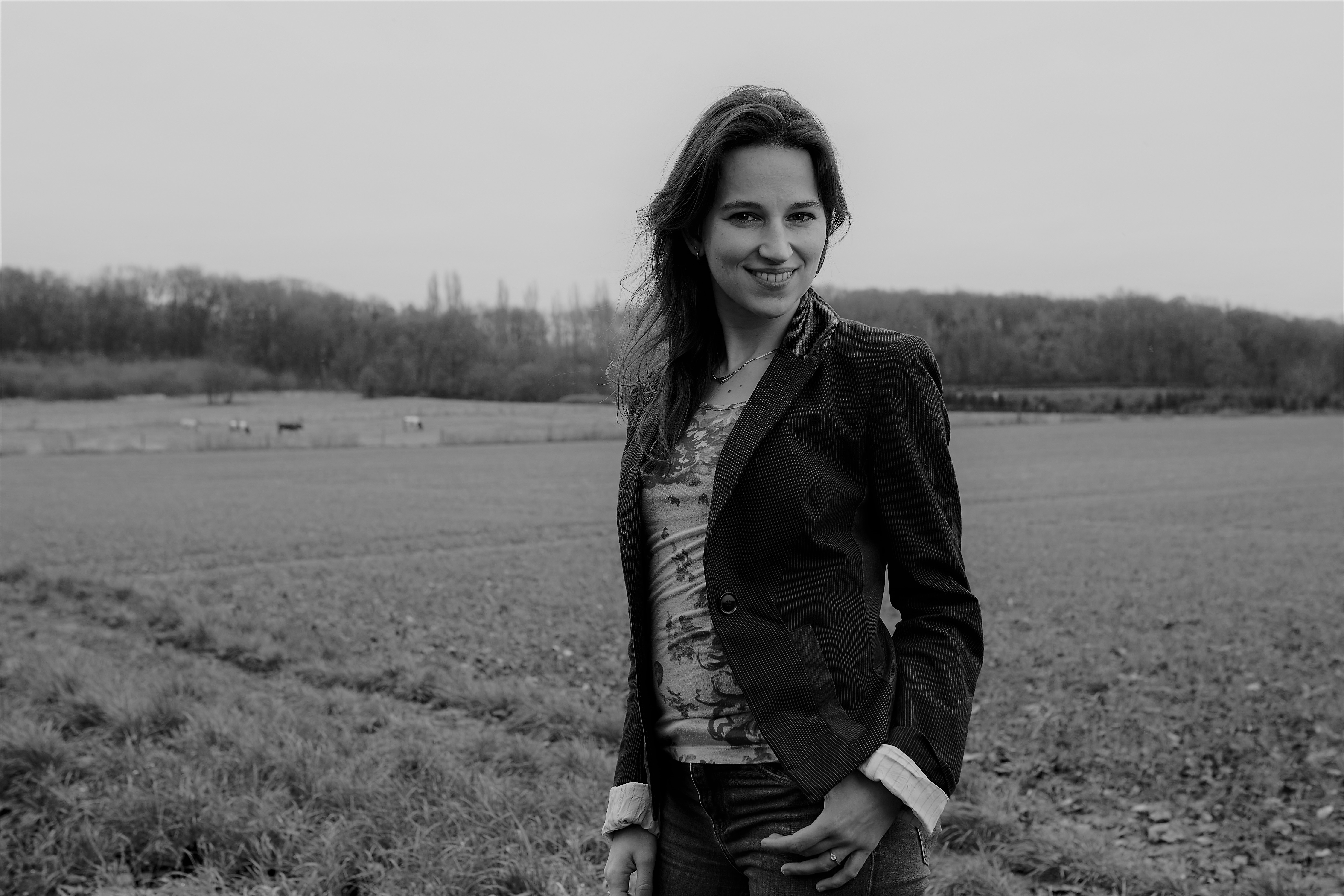}}]{Lianne Ippel}
		Lianne Ippel received her PhD in Statistics from Tilburg University on analyzing data streams with dependent observations, for which she won the dissertation award from General Online Research conference (2018). After a Postdoc at Maastricht University, she now works at Statistics Netherlands where she works at the methodology department on international collaborations and innovative methods for primary data collection.
	\end{IEEEbiography}
	\begin{IEEEbiography}[{\includegraphics[width=1in,height=1.25in,clip,keepaspectratio]{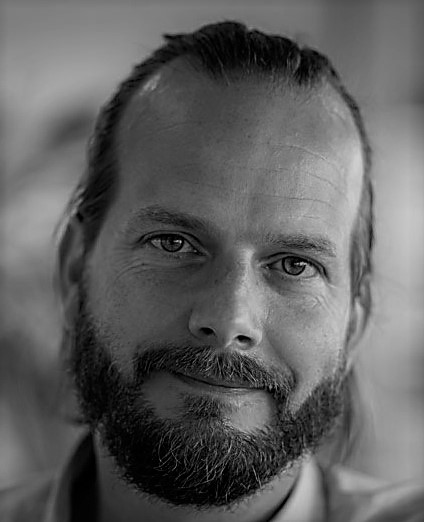}}]{Andre Dekker}
		Prof. Andre Dekker, PhD (1974) is a medical physicist and professor of Clinical Data Science at Maastricht University Medical Center and Maastro Clinic in The Netherlands. His Clinical Data Science research group (50 staff) focuses on 1) federated FAIR data infrastructures, 2) AI for health outcome prediction models and 3) applying AI to improve health. Prof. Dekker has authored over 200 publications, mentored more than 30 PhD students and holds multiple awards and patents on the topic of federated data and AI. He has held visiting scientist appointments at universities and companies in the UK, Australia, Italy, USA and Canada.
	\end{IEEEbiography}
	\begin{IEEEbiography}[{\includegraphics[width=1in,height=1.25in,clip,keepaspectratio]{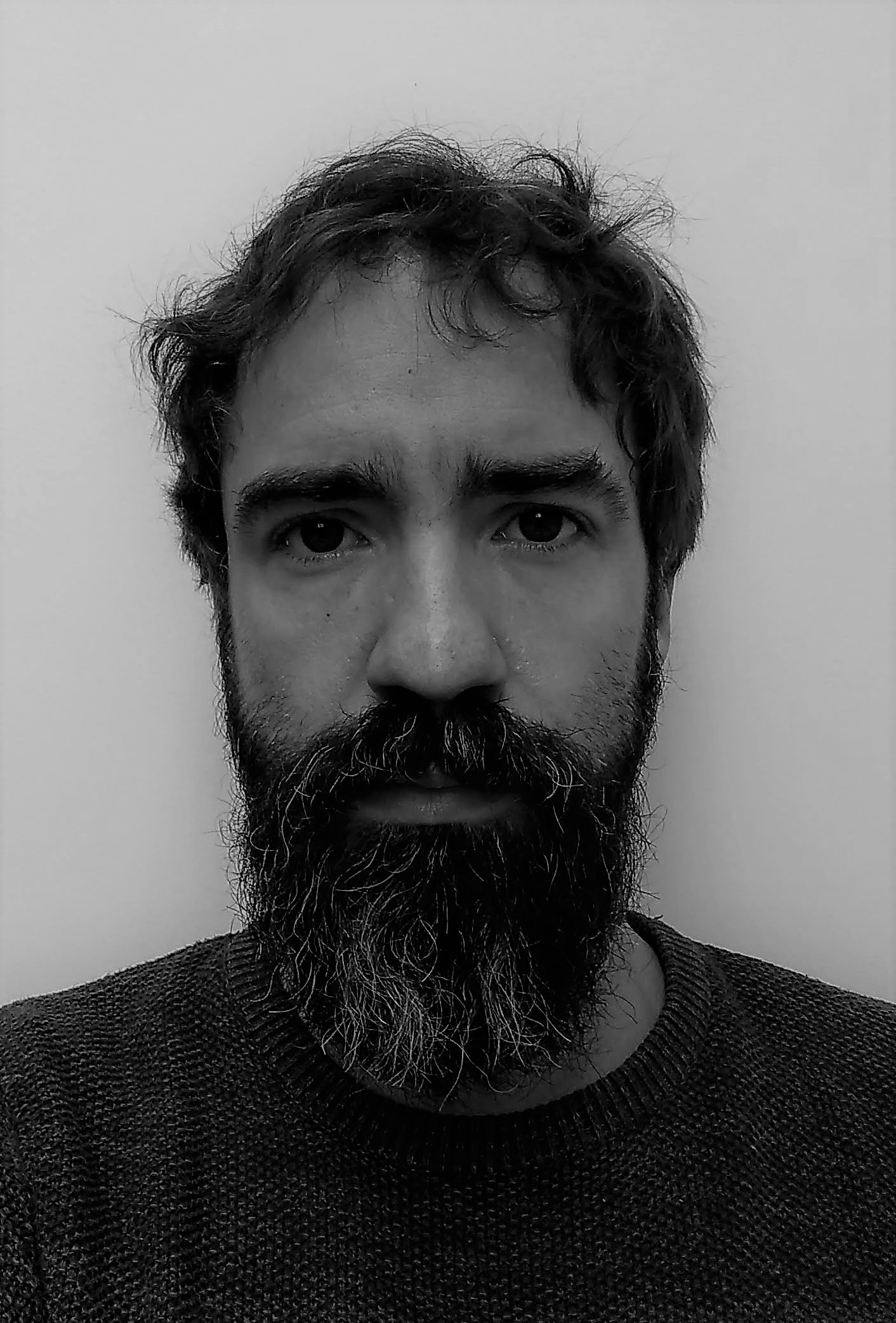}}]{Inigo Bermejo}
		Inigo Bermejo received the BSc degree on Computer Engineering from the University of the Basque Country, Spain, in 2006 and the PhD in Intelligent Systems from UNED, Spain, in 2015. He is currently a postdoctoral researcher at the Clinical Data Science group, Maastricht University. His research interests include privacy preserving techniques, prediction modelling and causal inference. 
	\end{IEEEbiography}

	\appendices
	\section{Full $3$-party naïve calculation}
	\label{Full 3-party naïve calculation}
	
	The full calculation can be expanded as follows:
	\setlength{\arraycolsep}{0.0em}
	\begin{gather*}
		\varphi(  \bold{\hat{A}} \cdot \bold{\hat{C}} \cdot \bold{B}){+}r_{b} {-} v_{2} {-} \varphi(\bold{R_{a}} \cdot   \bold{\hat{B}} \cdot \bold{\hat{C}}) {+} r_{a} \\
		{-}\: \varphi( \bold{R_{c}} \cdot \bold{\hat{A}} \cdot   \bold{\hat{B}}){+} r_{c} {+} v_{2}
		\\=\\ 
		\varphi(({\bf A+R_a})\cdot ({\bf C+R_c})\cdot {\bf B}) - \varphi(({\bf B+R_b})\cdot ({\bf C+R_c})\cdot {\bf R_a}) \\
		{-}\: \varphi(({\bf A+R_a})\cdot ({\bf B+R_b})\cdot {\bf R_c}) + r_a + r_b + r_c
		\\=\\
		\varphi(\bold{A} \cdot \bold{B} \cdot \bold{C} {+} \bold{A} \cdot \bold{B} \cdot 	\bold{R_{c}} {+} \bold{B} \cdot \bold{C} \cdot \bold{R_{a}} {+} \bold{B} \cdot \bold{R_{a}} \cdot \bold{R_{c}}) \\
		{-}\:\varphi(\bold{B} \cdot \bold{C} \cdot \bold{R_{a}} {+} \bold{B} \cdot \bold{R_{a}} \cdot \bold{R_{c}} {+} \bold{C} \cdot \bold{R_{a}} \cdot \bold{R_{b}} {+} \bold{R_{a}} \cdot \bold{R_{b}} \cdot \bold{R_{c}}) \\
		{-}\:\varphi( \bold{A} \cdot \bold{B} \cdot \bold{R_{c}} {+} \bold{A} \cdot \bold{R_{b}} \cdot \bold{R_{c}} {+} \bold{B} \cdot \bold{R_{a}} \cdot \bold{R_{c}} {+} \bold{R_{a}} \cdot \bold{R_{b}} \cdot \bold{R_{c}})
		\\{+}\: r_{a} {+} r_{b} {+} r_{c}
		\\=\\
		\varphi(\bold{A} \cdot \bold{B} \cdot \bold{C}) {+} \varphi(\bold{A} \cdot \bold{B} 	\cdot \bold{R_{c}}) {+}\varphi(\bold{B} \cdot \bold{C} \cdot \bold{R_{a}}) {+} \varphi(\bold{B} \cdot \bold{R_{a}} \cdot \bold{R_{c}}) \\
		{-}\:\varphi(\bold{B} \cdot \bold{C} \cdot \bold{R_{a}}) {-} \varphi(\bold{B} \cdot \bold{R_{a}} \cdot \bold{R_{c}}) {-} \varphi(\bold{C} \cdot \bold{R_{a}} \cdot \bold{R_{b}}) 
		\\{-}\:\varphi(\bold{R_{a}} \cdot \bold{R_{b}} \cdot \bold{R_{c}})
		{-} \varphi(\bold{A} \cdot \bold{B} \cdot \bold{R_{c}}) {-} \varphi(\bold{A} \cdot \bold{R_{b}} \cdot \bold{R_{c}})\\
		{-}\:\varphi(\bold{B} \cdot \bold{R_{a}} \cdot \bold{R_{c}}) {-} \varphi(\bold{R_{a}} \cdot \bold{R_{b}} \cdot \bold{R_{c}}) {+} r_{a} {+} r_{b} {+} r_{c}
		\\=\\
		\varphi(\bold{A}\cdot \bold{B} \cdot \bold{C}) {-} \varphi(\bold{C} \cdot 	\bold{R_{a}} \cdot \bold{R_{b}}) {-} \varphi(\bold{A} \cdot \bold{R_{b}} \cdot \bold{R_{c}})\\
		{-}\:\varphi({\bf B}\cdot {\bf R_a}\cdot {\bf R_c}) - \varphi({\bf R_a}\cdot {\bf R_b}\cdot {\bf R_c}) + r_a + r_b + r_c
	\end{gather*}
	
	\section{Full $3$-party example}
	\label{Full 3-party example}
	Practical example of the $n${-}party scalar protocol:
	$3$ parties Alice, Bob, \& Claire with the following data.
	Data $\bold{A}$: 
	$\begin{bmatrix}
		1 & 0  & 0\\ 
		0 & 1 & 0\\ 
		0 & 0 & 1
	\end{bmatrix}$
	Data $\bold{B}$: 
	$\begin{bmatrix}
		0 & 0  & 0\\ 
		0 & 1 & 0\\ 
		0 & 0 & 1
	\end{bmatrix}$ 
	Data $\bold{C}$:
	$\begin{bmatrix}
		1 & 0  & 0\\ 
		0 & 0 & 0\\ 
		0 & 0 & 1
	\end{bmatrix}$
	
	This means we are dealing with an $n${-}party protocol where $n = 3$. The target value would be: $\varphi (\bold{A} \cdot \bold{B} \cdot \bold{C}) = 1$
	
	Using the $n${-}scalar protocol the calculation will look as follows:
	First trusted third party Merlin generates the following three random matrices:
	
	$\bold{R_{a}}: 
	\begin{bmatrix}
		172 & 0  & 0\\ 
		0 & 243 & 0\\ 
		0 & 0 & 136
	\end{bmatrix} 
	\bold{R_{b}}:
	\begin{bmatrix}
		274 & 0  & 0\\ 
		0 & 356 & 0\\ 
		0 & 0 & 180
	\end{bmatrix} 
	\bold{R_{c}}: 
	\begin{bmatrix}
		341 & 0  & 0\\ 
		0 & 357 & 0\\ 
		0 & 0 & 69
	\end{bmatrix}$
	
	Merlin then calculates: $\varphi(\bold{R_{a}} \cdot \bold{R_{b}} \cdot \bold{R_{c}}) = 48643124$ Merlin then splits $\varphi(\bold{R_{a}} \cdot \bold{R_{b}} \cdot \bold{R_{c}})$ into three secret shares: $r_{a} = 8015322, r_{b} = 10543269$, \& $r_{c} = 30084533$. 
	
	Alice then calculates $\bold{\hat{A}} = \bold{A} {+} \bold{R_{a}} =
	\begin{bmatrix}
		173 & 0  & 0\\ 
		0 & 244 & 0\\ 
		0 & 0 & 137
	\end{bmatrix} $
	and shares the result with the others. Bob then calculates   $\bold{\hat{B}} =\bold{B}{+} \bold{R_{b}} = 
	\begin{bmatrix}
		274 & 0  & 0\\ 
		0 & 357 & 0\\ 
		0 & 0 & 181
	\end{bmatrix} $
	and shares the result with the others. Claire then calculates $\bold{\hat{C}} = \bold{C} {+} \bold{R_{c}} = 
	\begin{bmatrix}
		342 & 0  & 0\\ 
		0 & 357 & 0\\ 
		0 & 0 & 70
	\end{bmatrix} $
	and shares the result with the others. Alice generates a random value $v_{2} = 3$, after which Alice calculates:
	\begin{gather*}
		u_{1}
		\\=\\
		\bold{\hat{B}} \cdot \bold{\hat{C}} \cdot \bold{A} {+} (n{-}1) \cdot r_{a} {-} v_{2} 
		\\=\\  
		\varphi(  
		\begin{bmatrix}
			274 & 0  & 0\\ 
			0 & 357 & 0\\ 
			0 & 0 & 181
		\end{bmatrix} 
		\cdot 
		\begin{bmatrix}
			342 & 0  & 0\\ 
			0 & 357 & 0\\ 
			0 & 0 & 70
		\end{bmatrix} 
		\cdot 
		\begin{bmatrix}
			1 & 0  & 0\\ 
			0 & 1 & 0\\ 
			0 & 0 & 1
		\end{bmatrix}  ) {+} (3{-}1) \cdot 8015322 {-} 3
		\\=\\
		233827 {+} 16030644 {-} 3 
		\\=\\
		16264468
	\end{gather*}
	
	Bob then calculates
	
	\begin{gather*}
		u_{2}
		\\=\\
		u_{1} {-} \varphi(\bold{\hat{A}} \cdot \bold{\hat{C}} \cdot \bold{R_{b}}) {+} (n{-}1)r_{b} 
		\\=\\
		u_{1} {-} \varphi( 
		\begin{bmatrix}
			173 & 0  & 0\\ 
			0 & 244 & 0\\ 
			0 & 0 & 137
		\end{bmatrix} 
		\cdot 
		\begin{bmatrix}
			342 & 0  & 0\\ 
			0 & 357 & 0\\ 
			0 & 0 & 70
		\end{bmatrix} 
		\cdot 
		\begin{bmatrix}
			274 & 0  & 0\\ 
			0 & 356 & 0\\ 
			0 & 0 & 180
		\end{bmatrix} 
		\cdot ) \\
		{+}\:(3{-}1)\cdot 10543269 
		\\=\\
		16264468 {-} 48948132 {+} 21086538 
		\\=\\
		{-}11597126
	\end{gather*}
	
	Claire then calculates 
	\begin{gather*}
		u_{3}
		\\=\\
		u_{2} {-}\varphi( \bold{\hat{A}} \cdot   \bold{\hat{B}} \cdot \bold{R_{c}}) {+} (n{-}1) r_{c} 
		\\=\\
		u_{2} {-} \varphi( 
		\begin{bmatrix}
			173 & 0  & 0\\ 
			0 & 244 & 0\\ 
			0 & 0 & 137
		\end{bmatrix} 
		\cdot 
		\begin{bmatrix}
			274 & 0  & 0\\ 
			0 & 357 & 0\\ 
			0 & 0 & 181
		\end{bmatrix} 
		\cdot 
		\begin{bmatrix}
			341 & 0  & 0\\ 
			0 & 357 & 0\\ 
			0 & 0 & 69
		\end{bmatrix} ) \\
		{+} (3{-}1)\cdot 30084533
		\\=\\
		{-}11597126{-} 48972631{+}60169066
		\\=\\
		{-}400691
	\end{gather*}
	
	At this point $u_{3}$ is equal to the following: 
	\begin{gather*}
		u_{3} =	\varphi(\bold{A} \cdot \bold{B} \cdot \bold{C}){-} \varphi(\bold{A} \cdot \bold{R_{b}} \cdot \bold{R_{c}}) {-} \varphi(\bold{B} \cdot \bold{R_{a}} \cdot \bold{R_{c}}) \\
		{-}\:\varphi(\bold{C} \cdot \bold{R_{a}} \cdot \bold{R_{b}}) {-} v_{2}
	\end{gather*}
	
	The leftover terms in $\varphi(\bold{A} \cdot \bold{R_{b}} \cdot \bold{R_{c}}) {-} \varphi(\bold{B} \cdot \bold{R_{a}} \cdot \bold{R_{c}})\\ 
	{-}\:\varphi(\bold{C} \cdot \bold{R_{a}} \cdot \bold{R_{b}})$ need to be solved separately using their own $2${-}party scalar product protocol. Once these have been solved separately Claire calculates the following. For the sake of readability we introduce a helper variable $h$ here.
	
	\begin{gather*}
		h
		\\=\\ 
		u_{3} {+} \varphi(\bold{A} \cdot \bold{R_{b}} \cdot \bold{R_{c}}) {+} \varphi(\bold{B} \cdot \bold{R_{a}} \cdot \bold{R_{c}}) {+} \varphi( \bold{C} \cdot \bold{R_{a}} \cdot \bold{R_{b}}) 
		\\=\\
		u_{3}{+} \varphi( 
		\begin{bmatrix}
			1 & 0  & 0\\ 
			0 & 1 & 0\\ 
			0 & 0 & 1
		\end{bmatrix} 
		\cdot 
		\begin{bmatrix}
			274 & 0  & 0\\ 
			0 & 356 & 0\\ 
			0 & 0 & 180
		\end{bmatrix} 
		\cdot 
		\begin{bmatrix}
			341 & 0  & 0\\ 
			0 & 357 & 0\\ 
			0 & 0 & 69
		\end{bmatrix} ) {+} 
		\varphi(
		\begin{bmatrix}
			0 & 0  & 0\\ 
			0 & 1 & 0\\ 
			0 & 0 & 1
		\end{bmatrix} 
		\\
		\cdot\: 
		\begin{bmatrix}
			172 & 0  & 0\\ 
			0 & 243 & 0\\ 
			0 & 0 & 136
		\end{bmatrix} 
		\cdot 
		\begin{bmatrix}
			341 & 0  & 0\\ 
			0 & 357 & 0\\ 
			0 & 0 & 69
		\end{bmatrix} ) 
		\\{+}\:
		\varphi(
		\begin{bmatrix}
			1 & 0  & 0\\ 
			0 & 0 & 0\\ 
			0 & 0 & 1
		\end{bmatrix} 
		\cdot 
		\begin{bmatrix}
			172 & 0  & 0\\ 
			0 & 243 & 0\\ 
			0 & 0 & 136
		\end{bmatrix} 
		\cdot 
		\begin{bmatrix}
			274 & 0  & 0\\ 
			0 & 356 & 0\\ 
			0 & 0 & 180
		\end{bmatrix} )
		\\=\\
		{-}400691{+}232946{+}96135{+}71608
		\\=\\
		{-}2
	\end{gather*}
	
	Alice then calculates $h{+}v_{2} = {-}2 {+} 3 = 1$ which is our final result and corresponds to our expected result.
	
	\section{GIT repository}
	\label{git}
	An implementation of the $n$-party protocol in both java and in python can be found in the following git repo: \url{https://github.com/MaastrichtU-CDS/n-scalar-product-protocol}

\end{document}